# Circumstellar Habitable Zones to Ecodynamic Domains: A Preliminary Review and Suggested Future Directions

M. J. Heath

Ecospheres Project, 47 Tulsemere Road, London, SE27 9EH, UK

L. R. Doyle

SETI Institute, 515 N. Whisman Road, Mountain View, California 94043, USA

**Abstract.**

The concept of the Circumstellar Habitable Zone has served the scientific community well for some decades. It slips easily off the tongue, and it would be hard to replace. Recently, however, several workers have postulated types of habitable bodies which might exist outside the classic circumstellar habitable zone (HZ). These include not only bodies which orbit at substantial distances from their parent stars, but also snowball worlds with geothermally-maintained internal oceans and even densely-atmosphered worlds with geothermally-maintained surface oceans, which have been ejected from unstable planetary systems into interstellar space. If habitability is not a unique and diagnostic property of the HZ, then the value of the term has been compromised in a fundamental way. At the same time, it has become evident that multiple environmental states, differing in important ways in their habitability, are possible even for geophysically similar planets subject to similar levels of insolation, within the classic HZ. We discuss an approach to investigations of planetary habitability which focusses on planetary-scale ecosystems, which are here termed "*ecospheres.*" This is following a usage popular amongst ecologists, such as Huggett (1999), rather than that of authors such as Strughold (1953) and Dole (1964), who used it as a term for the HZ. This approach emphasises ecodynamic perspectives, which explore the dynamic interactions between the biotic and abiotic factors which together comprise ecosystems.

**Introduction: Habitability & Ecodynamics**

A meeting American Geophysical Union, Session B30: *Revisiting the Habitable Zone* (December, 2009, San Francisco, California) was announced with the statement: "*Traditionally the habitable zone has been defined as the region around a main sequence star in which terrestrial-like planets, with Earth-like atmospheres, can support surface water. This definition has served as an intellectual framework for interpreting potential habitability of exoplanets, but has not been significantly revised in the 16 years since the calculations of Kasting et al. (1993). With the first unambiguous discovery of terrestrial exoplanets made this year, as well as the recent launch of Kepler, there is renewed interest in determining additional constraints on, alternative routes to, and outstanding issues for planetary habitability. This session will explore various types of habitable zones, including, but not limited to, surface and subsurface habitable zones related to radiative, geophysical, and compositional effects. Selected abstracts will present new concepts, and will foster an interdisciplinary dialogue among geophysicists, atmospheric scientists, planetary scientists, and astronomers as the physical phenomena of habitability are modeled, and ultimately observed.*"

In answering the suggestion that the intellectual framework of the HZ "*…has not been significantly revised in the 16 years since the calculations of Kasting et al. (1993)*" it is necessary to distinguish between:

     1) Advances in areas of planetary science relevant to habitability;

     2) Concomitant advances in nomenclature and epistemology as regards the designation of the HZ.

The first interpretation of the statement about the HZ would clearly be untrue. It would be incorrect to assert that there have been few advances in branches of science pertinent to habitability or the HZ over the last 16 years. Indeed, as early as the beginning of 1994, the NASA-funded conference: *First International Conference on Circumstellar Habitable Zones* (January 19th to 21st, 1994) at the NASA Ames Research Center, Moffett Field, California (Doyle, 1996) was held to address just such issues. The idea behind the conference was to bring together scientists from a wide range of fields - including chemistry, geology, climatology, planetary science, astrophysics and biology – all which are of direct relevance to HZ research, but whose practitioners from different disciplines had too often failed to mutually communicate. Certain issues, such as the habitability of planets orbiting red dwarf stars were seriously addressed for the first time at that meeting (Haberle et al. 1996) and were pursued vigorously over the following years (eventuating in the NASA Astrobiology Institutes). At this time the first observational program capable, in principle, of determining the existence of terrestrial-sized planets by the transit method—and concentrating on the eclipsing red dwarf binary CM Draconis (Doyle *et al.*, 1996a; Doyle *et al.*, 2000; Deeg, *et al.*, 2000)—was begun. Over the next five years this effort set useful upper limits on planets around this system (to 2.3 Earth radii in the HZ) by applying signal detection methods (Jenkins et al. 1996) while demonstrating the utility of the transit method for edge-on double star systems (i.e., eclipsing binaries).

In fact, many workers—representing a wide range of academic institutions—have contributed key advances in planetary habitability studies and knowledge of the natural sciences has advanced on a broad front since 1993. Ongoing discoveries and paradigm revisions in astrophysics, planetary system dynamics, geology, biology and ecology are continuously driving changes in the way in which we investigate planetary habitability. This field, under-funded and under-publicised though it may be, is not a stagnant pool, but a fast-flowing stream. We are concerned here with the second interpretation, which requires more detailed discussion. It would probably be correct to say that there has been relatively little published specifically concerning the semantics and epistemology of the HZ concept. As Shock & Holland (2007) recognised (p. 839): "*The word habitability has entered the jargon of science without a precise definition.*" As HZ concepts have evolved, we found it more helpful to investigate ecodynamic domains (EDs), within which planetary-scale ecosystems (ecospheres in the parlance of ecologists; Huggett, 1998) could be structured and function in given modes, rather than classic HZs.

In their controversial book "*Rare Earth. Why Complex Life Is Uncommon in the Universe,*" Ward & Brownlee (2000) argued that the probability of any given planet having a suitable combination of environmental factors for the support of complex life was very low. They contended also that the HZ for higher life was probably narrow. They defined an "*Animal Habitable Zone.*" They did not apply climatic models to quantify the width of any of the HZs to which they referred, but it was considered that their animal HZ was wider than the zone in which human beings (together with their agriculture) could live, but narrower than that for higher plants. We consider that such an approach is fragmentary, and does not address the natural world as a systemic unity. Given that primary producers may (in principle) exist in the absence of heterotrophs, there would be more justification for adopting a photosynthetic HZ than the AHZ of Ward & Brownlee (2000). Howbeit, there are innumerable instances of plants and animals having co-evolved into mutual dependence. We contend that in the final analysis, organisms exist within ecosystems, and that they cannot be treated meaningfully in isolation from their ecosystems.

The term "*ecosystem*" was introduced by Tansley (1935, p. 299): "*There is a constant interchange within each system, not only between organisms but between the organic and the inorganic. These **ecosystems**, as we may call them, are of the most various kinds and sizes. They form one category of the multitudinous physical systems of the universe*". The ecosystem (p. 306) is: "*The fundamental concept appropriate to the biome considered together with all the effective inorganic factors of its environment*". An ecosystem thus includes not only organisms, but also (Tansley, 1939, p. 228): "*all the physical and chemical components*" of the associated environment. Ecosystems are thus not static, but are dynamic. They involve metabolism, population dynamics and and biogeochemical cycles. The dynamical style of a planetary scale ecosystem will be a key factor in determining habitability. A useful term was coined by Tricart & KiewietdeJonge (1992, p. 157): "*"Ecodynamics" is the study of*

*the ecological environment.*" It is "*concerned with the various processes and mechanisms that cause changes in the ecological environment. It studies how they function in space and time . . . it also examines their mutual interactions, conformable to a systems approach.*"

We have found it convenient to recognise:

a) Ecodynamic Regimes, each of which is distinguished by a given mode of ecosphere structure and function (with life able to modify its environment to various degrees);

b) Ecodynamic Phases in planetary history, each characterised by its own distinct ecodynamic regime; and

c) Ecodynamic Zones, which are volumes within astronomical space within which given Ecodynamic Regimes are viable.

Ecodynamic regimes, phases and zones occur within what may be denoted as Ecodynamic Domains in the phase space of environmental parameters. Outside of its domain, a given ecodynamic regime would not be viable. An ecosphere will survive as long as it can pursue a viable organisational trajectory through phase space connecting the origin of life (be it *in situ* or arrival by panspermia) and the specified state at a later stage. This conceptual framework, we would suggest, has the advantage of not having begun its working life tied to the implication that Ecodynamic Zones (EZs) are of necessity circumstellar, even if *some* EZs must be obligatorily circumstellar.

This defintion is also flexible so that different researchers are free to define specific regimes, phases, zones or domains as their own particular purposes require. These concepts can be used alongside existing terminology relating to the HZ, or the established geological periods endorsed by the International Commission on Stratigraphy. Geological periods are not characterised by unchanging conditions throughout their duration, and sometimes, one may wish to follow given trends constituting identifiable ecodynamic phases (such as, for example, oxygenation, rises and declines in reconstructed biodiversity, or the assembly or break-up of supercontinents) across the boundaries of conventional periods.

The impetus to devise a different terminology and another conceptual framework came from the fact that the HZ concept, whilst remaining useful for some purposes, fails to accommodate certain classes of habitable bodies which have been hypothesised. Its shortcomings cannot be overcome in every instance by adding suffixes or prefixes to subdivide or expand it, whilst retaining the core designation of "*Habitable Zone.*" In order to be meaningful at all, the term "*Habitable Zone*" must refer to a clearly defined zone, whose unique and diagnostic characteristic is that it is habitable. If habitability lies outside the HZ, or if it is not confined to a discrete zone, then the term can have no real scientific value. We consider first modifications of the HZ concept which would leave it essentially intact.

Kasting *et al.* (1993) was a definitive landmark, and it set the agenda for much future work. It explored the HZ in relation to climate models which set limits on the ability of an Earth-like planet to retain liquid water on its surface at higher and lower levels of insolation that are incident upon the Earth ($I_\oplus$) at the present day. The basis of these models was the carbonate-silicate cycle, which may act as a rough planetary thermostat, dampening fluctuations in mean planetary surface temperature ($T_{surf}$). (Walker *et al.*, 1981). Were insolation lower, weathering rates would adjust so that there would be a higher atmospheric partial pressure of $CO_2$ ($pCO_2$). Were insolation higher, $pCO_2$ would fall. This process offers an explanation of how the early Earth largely avoided freezing over, when the Sun was significantly fainter than today (although bacterially-produced $CH_4$ could have figured also; for example Pavlov *et al.*, 2000 – which would permit lower values of $pCO_2$). Williams *et al.* (1997) investigated the potential habitability of giant planetary moons, arguing that a body of mass $\geq 0.23\ M_\oplus$ (about twice that of Mars) should be able to retain an atmosphere and (ignoring the possibility of tidal heating) sustain sufficient (radiogenically produced) geological activity to enable a carbonate-silicate cycle for several Gyr.

Kasting *et al.* put forward a number of case studies, but their standard case assumed that they were dealing with an Earth-like planet which possessed an ocean of 1.4 x $10^{21}$ litres of water, and that there was a background 1.0 bar $N_2$ atmosphere (although other cases were considered also in their paper) and their analysis followed the one-dimensional radiative-convective climate model which had been explored in Kasting *et al.* (1984) and Kasting & Ackerman (1986).

Their conservative estimate for the inner margin of the HZ around the present day Sun (based on Rassool & DeBergh, 1970) was ~ 0.95 AU, corresponding to 1.1 $I_≈$ (they adopted a value of 1,360 $Wm^{-2}$ for 1.0 $I_≈$). Here, slow water loss (moist greenhouse conditions) would prevail. Runaway greenhouse conditions would begin at 1.4 $I_≈$ (~ 0.85 AU from the present Sun); the oceans would evaporate into the atmosphere, giving a pressure of 270 bars, and water loss would be accelerated. This was known as the "*runaway greenhouse*" condition. Under runaway greenhouse conditions, Earth's oceans could have been lost on a time-scale of < 30 Myr.

The outer margin of the HZ was defined as the radial distance from a star at which the condition $T_{surf} ≤ 0°C$ would apply. $CO_2$ is significantly more effective at back-scattering solar radiation back into space than normal Earth air. Consequently, there will come a point of "*maximum greenhouse*" limit, at which the reflectivity (albedo) of a planet is sufficiently high to negate the greenhouse effect from rising $pCO_2$. On this basis, the outer margin of the HZ was placed at ~ 1.67 AU (0.35 $I_≈$). However, beginning at ~ 1.37 AU (0.53 $I_≈$), $CO_2$ ice clouds would begin to condense in the atmosphere, and they also would reduce the amount of energy absorbed by a planet.

In contrast, Forget & Pierrehumbert (1997) argued that down-scattering of IR by $CO_2$ clouds of ice crystals might actually extend the HZ outwards significantly. According to their study, a planet could maintain $T_{surf} > 0°C$ at 2.4 A.U. from the Sun (with 0.17 $I_≈$), provided that it possessed an atmosphere with $pCO_2$ = 10 bars.

A key concept, the "*Continuously Habitable Zone*," was introduced by Hart (1978; 1979), as a zone within which a planet could sustain life-supporting conditions for evolutionary time-scales despite the increase in luminosity of its parent star during main sequence evolution and the associated outward migration of HZ boundaries. Kasting *et al.* (1993, p. 108): "*A conservative estimate for the width of the 4.6-Gyr continuously habitable zone (CHZ) is 0.95 to 1.15 AU.*" The CHZ concept emerged naturally from the climatic models for the HZ, and it extends the concept without introducing conceptual or semantic problems.

**Extension of the HZ into the Vicinity of M Dwarfs.**

A natural sub-division of the HZ, again, conceptually unproblematic, involves that section of the HZ which lies sufficiently close to a star for a planet to become tidally-locked into synchronous rotation. It would exist for M-dwarf stars, whose faintness requires any planet receiving insolation (at all wavelengths) comparable to 1.0 $I_≈$ to be orbiting close in and thereby subject to intense tidal forces. Such planets will suffer rotational capture (become tidally locked in synchronous rotation with their orbital periods) because, while the circumstellar HZ distance scales as the inverse of the square-root of the stellar luminosity, the gravitational tidal forces scale as the inverse cube-root of the stellar mass. Kasting *et al.* (1993) considered (p. 125): "*one should not rule out M stars altogether as possible abodes for life . . .*" However, "*All things considered, M stars rank well below G and K stars in their potential for harboring habitable planets.*"

Faint red dwarfs (spectral class M V)—which constitute greater-than 70 % of all stars in the galaxy—(Zelik, 2002) were essentially previously excluded from discussion of the HZ. There was a widespread opinion that both atmosphere and oceans could freeze out on the permanently dark hemisphere of a planet locked into synchronous rotation. However, Haberle *et al.*, (1996) and Joshi *et al.*, (1997) were able to demonstrate that even atmospheres resembling that of the present day Earth would carry sufficient heat to the unlit hemisphere to prevent atmospheric freeze-out. Heath *et al.* (1999) contributed the conclusion that a geothermal flux similar to that of the present day Earth would maintain dark hemisphere sea ice at an equilibrium thickness. Snowfall arriving at the surface of the dark side sea ice would be balanced by melting at the base of the sea ice. Given Earth-like ocean basin geometry—with melt water entering seas which communicated with those on the lit hemisphere—there could be vigorous hydrothermal exchange between the hemispheres. Moreover, seas on the dark hemisphere did not freeze in all simulations. It was shown that Photosynthetically Active Radiation (PAR) could also be adequate for higher plant life, at least around the hottest M dwarfs, despite blanketing of the PAR region of the spectrum (particularly by titanium oxide absorption features). Also it was found that flare events need not repeatedly sterilise planetary surfaces. Results of improved climate simulations were published in Joshi (2003). Heath & Doyle (2004) explored the sequence of events on planets that span down over an extended geological time-scale through non-synchronous and near-synchronous into synchronous rotation.

In July 2005, NASA sponsored a *Workshop on the Habitability of Planets Orbiting M Stars* at the SETI Institute, whose attendees re-examined the problems from first principles upwards (which provided a valuable peer review), and endorsed similar conclusions to these previous papers in Tarter *et al.* (2007). Heath & Doyle (2010) also suggested how planets orbiting red dwarfs might obtain plentiful PAR, if their primaries were the binary companions to hotter stars, like the Sun. It remains to be shown observationally if such an arrangement is dynamically conducive to widespread planet formation. Kiang *et al.* (2007a; 2007b; 2008) published seminal papers, drawing on knowledge of the photosynthetic metabolism of plants on Earth, and discussing how photosynthesis might operate on planets receiving sunlight with a different spectral profile to that radiated by our own Sun.

Buccino *et al.* (2007) discussed a "*UV habitable zone*," and considered the problems of lack of UV in quiescent M-dwarf star output for forming molecules essential for initiating life, and the problems of strong UV flares in tandem. They concluded that UV around inactive M stars would be inadequate to synthesise complex macromolecules, whilst moderate flare activity could be beneficial. Since 2005, serious concerns have been expressed about the atmospheres of close-in planets of M dwarfs being blown away by the fierce winds of charged particles emitted by their suns (for example, Khodachenko *et al.*, 2007; Scalo *et al.*, 2007). Larger planets might best be able to generate their own protective magnetic fields, but simulations by Raymond *et al.* (2007) found that it could be difficult to form planets with > 0.3 Earth masses ($M_e$) around stars smaller than 0.5 - 0.8 solar masses ($M_s$).

Interestingly, Mayor *et al.* (2009) demonstrated that the minimum masses of the planets inferred for the system of Gliese 581 (a M3 dwarf of 0.32 $M_s$ and 0.013 solar luminosity) are 1.94 $M_e$, 15.65 $M_e$, 5.36 $M_e$ and 7.09 $M_e$ and analysis of the dynamics of the system constrains their masses to a maximum 1.6 times the minimum (M sin i) masses (where i is the planetary orbital inclination). The term "*super-Earth*" has been used by various authors to describe these bodies, but how these planets actually compare in composition and mode of formation with the Sun's terrestrial planets is not yet known. A paradigm shift was launched in 1994, but the implications are as yet uncertain. Whether M-dwarf stars may possess habitable planets must now be tested by observation. In any event, the potential extension of the HZ into a region close to M-dwarf stars would not constitute a major epistemological challenge to the classical HZ.

**Expansion of the HZ with Stellar Evolution**

The HZ boundaries must expand outwards with time as a star grows brighter during its main sequence evolution (because of the increasing density of post-fusion material in the core), and then leap outwards as the star enters into its red giant phase. A number of authors have addressed the issue, including Lorenz *et al.* (1997), who examined the possible future of Titan, a world on which life might develop; (p. 2,905): "*we find a window of several hundred Myr exists, roughly 6 Gyr from now, when liquid water-ammonia can form oceans on the surface and react with the abundant organic compounds there. . . . The duration of such a window exceeds the time necessary for life to have begun on Earth.*"

Stern (2003) defined a "*Delayed Gratification Habitable Zone*" in the outer Solar System, which would exist several Gyr from now as the Sun expanded into a luminous red giant. although the ability of icy worlds, such as we see in the Kuiper Belt, to sustain life once surface melting gets under way needs to be examined. Lopez *et al.* (2005) discussed the fate of the habitable zone around evolved stars of initial masses 1.0, 1.5 and 2.0 $M_s$, from the point at which a star moves off the main sequence until it reaches the point of the helium flash. von Bloh *et al.* (2009) modelled the case of a 10 $M_e$ super-Earth around a star that was evolving through its red giant phase, concluding (p. 593): "*We obtain a large set of solutions consistent with the principal possibility of life. The highest likelihood of habitability is found for ''water worlds.'' Only mass-rich water worlds are able to realize [post] HZ-type habitability beyond the stellar main sequence on the Red Giant Branch.*"

These investigations are not problematic in terms of the classic HZ, because the Delayed Gratification Habitable Zone is essentially a classic HZ displaced outwards to a much greater distance than during the star's main sequence phase. Although the climatology and biology of icy worlds with surface features brought to their melting points will not be directly comparable with those of terrestrial-type bodies with partial, or shallow ocean cover, they present us with a variant on the familiar theme of the HZ as previously defined.

**Habitable Bodies Outside the Classic HZ**

The situations outlined above would require non-challenging extensions or subdivisions of the HZ concept. Since 1993, however, when Kasting *et al.* published their milestone paper, scientists involved in planetary habitability studies have postulated several potential habitats for life, well outside the classic HZ, receiving very low quantities of insolation. Such objects do undermine the classical concept of the HZ in fundamental terms.

The implication of these developments has been noticed. In a review in *Science* entitled "*Expanding the Habitable Zone,*" Vogel (1999, p. 70) summarised new work that was challenging traditional ideas on the width of the HZ: "*Once restricted to a relatively narrow slice of the solar system, the possible environments for life in space are multiplying, reaching Pluto and even into interstellar space.*" She contended that "*for a star like the sun, traditional estimates extend from an orbit as close as Venus's (about 0.7 times Earth's orbit) to one just inside Mars (about 1.4 times Earth's orbit). But, if some other energy source can keep water liquid, life could flourish without a sun, and such habitable zone estimates would be way off.*" If these possibilities were to be accepted, the implication would be that the classical concept of the HZ was not so much expanded as abolished (English, 2000).

It should be borne in mind that whilst there are many hypothesized situations in which life could exist, we have as yet no observational confirmation of life beyond the Earth. On the other hand, however, if a hypothesis were too optimistic in the case of any given Solar System body, it might—with certain arbitrary modifications—be re-applied to classes of bodies which may exist around other stars. In a Galaxy with hundreds of billions of stars, it would be very hard to naysay a concept on the grounds of special pleading, because the special conditions required to make it work may be realised somewhere with so many circumstances possible.

Bodies which might host life—despite lying outside the classic HZ—include ice-bound objects with internal oceans whose outer shells are sufficiently thin (at least in places) for organisms to carry out photosynthesis in the higher plant PAR range (4,000 Å to 7,000 Å) beneath the ice as well as other ice-bound objects where light cannot penetrate and where ecospheres would thus have to be supported by other means.

We denote internal oceans as "*endo-oceans,*" as opposed to the surficial "*exo-ocean*" of the Earth. Water is the dominant component of Solar System icy bodies, and cosmochemical considerations make it very probable that such bodies are commonplace around other stars.

Ice-encased worlds are conveniently described as *"Snowball Worlds,"* a term used by Tajika (2008), who investigated their ability to host geothermally-sustained endo-oceans. They are a class of object whose existence appears to have been confirmed reliably for the Solar System. Khurana *et al.* (1998) reported perturbations in the Jovian magnetosphere in the vicinity of Europa and Callisto, measured by the Galileo spacecraft, which implied the existence of (p. 777): *"significant electrical conductivity just beneath the surfaces of both moons."* An explanation of how an endo-ocean could have survived within Callisto down to the present day was provided in a key study by Ruiz (2001), who reported (p. 409) that: *"previous work indicated that an outer ice layer on the ocean would be unstable against solid-state convection, which once begun would lead to total freezing of liquid water in about $10^8$ years. Here I show that when a methodology for more physically reasonable water ice viscosities (that is, stress-dependent nonnewtonian viscosities, rather than the stress-independent newtonian viscosities considered previously) is adopted, the outer ice shell becomes stable against convection."* A reviewer remarked (Bennett, 2001, p. 396): *"And of course, where there is any suggestion of water elsewhere in the Solar System other than Earth, there are suggestions of the possible existence of life."*

Ruiz' model did not require the presence of substances which could lower the freezing point of the primarily $H_2O$ ocean, but it was noted that $NH_3$ could reduce the melting point to as low as 176 K (Kargel, 1992).

For a surface temperature of 130 K and a present geothermal flux at the surface of 3.6 mWm$^{-2}$ (a compromise between 3.3 mWm$^{-2}$ for an undifferentiated Callisto and 3.9 mWm$^{-2}$ for a fully differentiated Callisto), the thickness of the outer icy shell would be < 105 km. In the deep interior of Callisto and other icy moons with an ice/ocean mass of sufficient depth, ice-I would convert to a high pressure, high density form of ice, at a depth determined by ambient temperature and pressure. Jupiter's other large icy moon, Ganymede, the largest moon in the Solar System, and Titan, Saturn's largest moon, and the second largest moon in the Solar System, are likewise among Solar System bodies of considerable interest in terms of their endo-oceans.

Much smaller bodies may possess internal quantities of water, although the term *"oceans"* may be over-enthusiastic in these cases. The discovery of a huge plume of dominantly $H_2O$, fed by multiple jets emerging from the ~ 500 km diameter Saturnian moon Enceladus (Porco *et al.*, 2006) has posed the question of whether this small, tidally-heated body possesses liquid water in its interior as well. If it does, then the possibility of life is wide open. Of course, the interior of Enceladus might contain bodies of liquid much colder than Earth's seas; $NH_3$, methanol and salts could reduce the freezing point of water to 176 K. The possibility of life having arisen on Enceladus was discussed by McKay *et al.* (2008). Not all workers have opted for an optimistic model involving internal bodies of water. Kieffer *et al.* (2006) argued that active tectonism was creating fractures in the south polar region, which was responsible for explosive de-gassing of clathrate. Waite *et al.* (2009), however, reported that $NH_3$ (with a mixing ration of $8.2 \pm 0.2 \times 10^{-3}$) has been detected in plume ejecta, together with organic compounds, deuterium and, very probably, $^{40}Ar$ (volume mixing ratio $3 \times 10^{-4}$). They concluded (p. 487): *"Ammonia's presence in the plume, along with the detection of Na and K salts in E-ring ice particles, implies that the interior of Enceladus may contain some amount of liquid water."*

Parkinson *et al.* (2008, p. 355) stated: *"We hypothesize that Enceladus' plume, tectonic processes, and possible liquid water ocean may create a complete and sustainable geochemical cycle that may allow it to support life. . . ."* Meyer and Wisdom (2007) thought that such a feature would have to be the aftermath of past orbital changes or episodic heat release.

Tidal heating is not necessarily required for small bodies to contain substantial amounts of internal water. Ceres (diameter ~ 950 km)—largest of the minor planets in the main asteroid belt between Mars and Jupiter—is a lone object with an orbital semi-major axis of 2.77 AU. A recent study of its geophysical evolution (Castillo-Rogez *et al.*, 2007)—taking into account short-lived radionuclides—suggests that complex differentiation is possible in this body. Beneath an outer icy shell, could be a layered core, with a small metallic centre, surrounded by dehydrated silicates, and an outer layer of hydrated silicates. Hydrothermal activity will have occurred, and may continue today. The existence of an endo-ocean was not ruled out, particularly if $NH_3$ were present.

Much smaller icy bodies than Ceres or Enceladus might even enjoy episodes during which they possessed internal bodies of water. For example, Wallis (1980) argued (p. 431): "*Comets accreted soon after the initial collapse and cooling of the solar nebula, and containing a plausible fraction of $^{26}Al$, would have been significantly heated as this radionuclide decayed. Snow-and-dust balls as described by integrals of the heat conduction equation would melt in the centre if larger than 3-6 km radius. A central, low pressure vapour-droplet mixture is described here, which is conceived to be retained within an ice shell, and providing a potentially habitable environment for elementary life forms. Refreezing after some million years produces a partially-hollow core.*" A hollow core, he contended, offered an explanation for the observed splitting of certain comets into fragments.

Photosynthetic production is vigorous at the surface of the Earth, and is the defining feature of terrestrial ecology. On some snowball worlds, even though they may be located outside the classic HZ, water-splitting photosynthesis may be possible because there are windows of thinner ice which admit light into the uppermost parts of endo-oceans. Such a model was advanced for Jupiter's icy moon Europa, following the passage in 1979 of Voyagers 1 and 2 through the Jupiter system. It was evident that Europa showed no signs of an early heavy bombardment, which was consistent with ongoing re-surfacing enabled by an elevated geothermal flux resulting from tidal interactions between Europa, Jupiter and the other Galilean moons. The Europa model of Squyres *et al.* (1983) involved an ocean at least tens of kilometres deep overlying the silicate body of the moon, and this was itself overlain by an icy shell averaging no more than 10 km thick. Adopting this model, Reynolds *et al.* (1983) postulated that photosynthetic life could exploit transient areas roofed with thinner, transparent or translucent ice cover. Opinions are polarised about whether the shell is tens of kilometres thick, with little opportunity for exchange of material from depth with surface materials (Schenk, 2002), or whether its thickness, at least locally, may be mere hundreds of metres. Greenberg *et al.* (2000) envisaged brines regularly reaching the surface. The dispute has been dubbed the "*great thickness debate*" in a discussion by Billings & Kattenhorn (2005).

We ask whether certain smaller worlds might have offered at least transient opportunities for photosynthesis. *If* windows of thinner ice existed on a cometary-type body, and *if* the proportion of dusty material was sufficient to provide nutrients, and if the body was sufficiently large to retain internal heat for long enough (there is a wide range of sizes of objects from the Kuiper Belt) then a relatively small sea inside a comet might indeed be a suitable home for life. Of interest is a calculation by Miller & Lazcano (1996) that the genetic complement of cyanobacteria may have taken less than 10 Myr to evolve.

There are several possibilities for supporting life on snowball worlds where PAR will not be available. One possibility is chemosynthesis involving reactions between water and minerals, which has been suggested as a means of supporting the bacterial ecosystems discovered on Earth in boreholes to depths of kilometres. Gold (1992) postulated that a major biomass of microbes existed at substantial depths in the crust. Stevens & McKinley (1995) claimed to have identified "*SLiMES*" (Subsurface Lithoautotrophic Microbial Ecosystems) living deep in the Columbia River Basalt Group of the Snake River Plain. They contended that bacteria could thrive on $H_2$ released in reaction between ground water and ferrous iron (review in Kaiser, 1995), although Anderson *et al.* (1998) were unable to confirm this process at environmentally relevant pH levels. Freund *et al.* (2002) noted that although $H_2$ can be liberated through reaction between water and fresh mineral surfaces, which involves the oxidation of ferrous iron, that (p. 83): "*A more reliable and potentially more voluminous $H_2$ source exists in nominally anhydrous minerals of igneous and metamorphic rocks.*" Lin *et al.* (2009) investigated the diversity and metabolic activity of microbial communities in a 3 to 4 km deep fracture through the 2.7 Gyr old Ventersdorp Supergroup metabasalt (Mponeng gold mine, South Africa). Lin *et al.* (2009, p. 482) wrote: "*the deep crustal biosphere may be energy-rich, is not approaching entropic death, and is capable of sustaining microbial communities indefinitely by geological processes.*"

Frederickson & Onstatt (1996) assumed that microbial life in general could extend down to the 110°C isotherm, with thermal gradients of 25°C km$^{-1}$ for continental and 15°C km$^{-1}$ for oceanic crust. Hyperthermophiles have been cultured at 113°C (Stetter, 1996). Pedersen (2000) observed (p. 13): "*The repeated observations of autotrophic, hydrogen-dependent microorganisms in the deep aquifers imply that hydrogen may be an important electron and energy source and carbon dioxide an important carbon source in deep subsurface ecosystems.*"

As Heath & Doyle (1996, and references therein) pointed out, deep sub-surface, geothermally-supported ecosystems could have provided places where life might have survived on the early Earth during episodes of intense bombardment, and they might well thrive on planets at arbitrary distances from stars. Vance *et al.* (2007) investigated how hydrothermal systems might operate on planets and satellites of < 1.0 $M_e$.

Irwin & Schulze-Makuch (2003) modelled Europa with two possible types of ecosystems. The first was a benthic chemosynthetic ecosystem, whilst the second possibility involved a novel mechanism. "*One producer is assumed to reside in the benthic zone, where ongoing dissolution of the mantle generates a hypertonic environment (one in which the solute concentration of the environment exceeds that of the organism's interior). This favors inward diffusion of solutes into the organism. At the top of the water column, melting ice dilutes the habitat to the point where osmotic pressure favors the inward movement of water. In both cases the transmembrane fluxes promoted by the ionic and osmotic gradients could be coupled to the formation of high-energy chemical bonds.*"

**HZs That Are Not Spherically Symmetrical.**

Most discussions of HZs assume that they will be described as spherical shells centred on a star. Other possibilities exist. Chyba (2000a; revised in 2000b) advanced a model in which (p. 381): "*disequilibrium chemistry in the ocean's ice cover, driven by charged particles accelerated in Jupiter's magnetosphere, should produce enough organic and oxidant molecules to fuel a substantial Europan biosphere.*" Hand *et al.* (2006) re-examined the question of how the surface of Europa is modified chemically by both high energy charged particles from the jovian magnetosphere and solar UV, and how products could be carried downwards. Formation of clathrates could modify the physical properties of the icy shell, causing it to become thinner.

Further problems are posed by habitable bodies which orbit stars, but which challenge the traditional HZ concentric shell geometry. For example, habitable conditions with classic $H_2O$ exo-oceans may prevail on planets in eccentric orbits which dip through the HZ (Williams & Pollard, 2002). As regards snowball worlds with endo-oceans, the HZ for biology—which depended upon synthesis of biologically useful chemicals through charged particles impacting on the surface of a snowball world from the magnetosphere of a giant planet—would not be spherically symmetrical, because of the trajectory along which particles must travel from stellar atmospheres where magnetic field lines are open, moving outwards in a warped current sheet and becoming concentrated in the magnetosphere. Stellar winds may have other impacts on planetary habitability. For example Chassefrière (1997) postulated that an early solar wind about $10^3$–$10^4$ more intense than at the present day and a solar UV flux around 5 times greater than today could have facilitated loss of water from Venus during the first few $10^8$ years of the Solar System. Doyle et al. (1996b; see also Whitmire et al. 1995) observed young solar-type stars at radio wavelengths to see if they were losing mass at rates sufficient to explain early water loss on Mars (i.e., the faint Sun paradox).

**Habitable Bodies in Interstellar Space**

The possibility of habitable bodies detached from stars into interstellar space, and orbiting the Galaxy as independent bodies, undermines the classical concept of the HZ concept entirely. Tajika's useful paper about snowball worlds (Tajika, 2008) proceeded from a recognition that (p. L54): "*The incident flux from the central star affects the surface temperature, hence the ice thickness. The*

*effect is estimated for different values of luminosity of the central star and the semimajor axis of the planetary orbit . . . . The results suggest that the ice thickness increases with distance from the central star because of a decrease in the incident flux. Considered how ice thickness could vary on oceanic terrestrial and super-terrestrial planets scaling geothermal flux with planetary mass.*"

An independent object of mass 0.4 $M_e$ was deemed unable to sustain an endo-ocean, for an appropriate duration, while a glaciated Earth-mass body could maintain an endo-ocean at around 4.0 AU from the present-day Sun. An object of mass at least 3.5 $M_e$ could, similarly, maintain an endo-ocean in the inner edge of the Kuiper Belt at 40 AU. Moreover, the internal ocean would be unable to freeze completely for planets with masses $\geq 4.0\ M_e$ even if the surface temperature were reduced to 0 K (assuming Earth-like characteristics—radiogenic heating, etc.)

Bearing that last figure in mind, Tajika's conclusion (p. L55): "*It is, therefore, suggested that if the subsurface ocean on the snowball planet could be habitable for life, life might well exist far outside the HZ for the traditional ocean planet*") is something of an understatement.

Pena-Cabrera & Durand-Manterola (2004) identified a "*Cold Circumstellar Habitable Zone*" in which $H_2O$ endo-oceans could occur on snowball worlds. However, as shown by Tajika (2008) there is no obvious outer margin of this zone; planets ejected from unstable systems into interstellar space could sustain endo-oceans.

It is not only bodies with endo-oceans which might be found in interstellar space. Stevenson (1999) pointed out that many models of planet formation would permit solid bodies of several Earth masses to accrete before the $H_2$ component of the early circumstellar nebula had dissipated. If such a planet were ejected from an unstable system into interstellar space, an Earth-like heat-flow of $4 \times 10^{20}$ erg s$^{-1}$—due only to the decay of long-lived radionuclides—could sustain an exo-ocean beneath a ~ 1.0 kbar $H_2$ atmosphere due to atmospheric opacity in the far-IR. Debes & Sigurdsson (2007) discussed how ejected binary planets could benefit from such tidal heating.

Bodies distant from their parent stars, and even interstellar bodies ejected from planetary systems, need not even be denied all sources of photosynthesis. Discussing the origins of photosynthesis, Björn (1995) discussed the possible energetics of IR photosynthesis using thermal emissions from sea floor hot water vents (see Vance *et al.*, 2007 for a discussion of how hydrothermal systems might operate on planets and satellites of < 1.0 $M_e$). The concept is not mere hypothesis; Beatty *et al.* (2005) have described what appears to be an obligately photosynthetic bacterial anaerobe from a deep-sea hydrothermal vent.

As demonstrated by Stevenson (1999), there is no outer margin for the surface liquid water HZ. Even an object ejected into interstellar space, could, if fortunate enough to possess a suitable atmosphere, maintain $T_{surf} \geq 0°C$ by trapping geothermal heat. For such bodies, and massive snowball worlds, the only uninhabitable places in the universe would lie in the close proximity of stars, or in the presence of violent astrophysical events.

The "*HZ*" is often introduced with the definite article. It is evident, however, that as it stands, the designation is a counter-truism because "*Habitable*" does not specify a unique and diagnostic characteristic of the zone which has been classically designated as habitable.

Moreover, whilst the term "*zone*" is correct only insomuch as habitable bodies must occur within a volume of astronomical space, it is obvious that planets which have escaped into interstellar space (or even, under some circumstances, been ejected from galaxies into intergalactic space) would not occupy a discrete zone in the sense that the HZ has previously been understood. The possibility of "Stevenson" worlds with exo-oceans and large interstellar snowball worlds with endo-oceans, cannot be discussed meaningfully within an expanded HZ concept. It is possible, however, to identify an Ecodynamic Domain in phase space within which ecospheres functioning like these are viable. It is possible also, since the term has no prior history—and hence no long-understood connotation of being centred upon a star—to define an Ecodynamic Zone in astronomical space where such ecospheres could occur, albeit, such a zone fills most of the observable universe.

Finally, the notion of clay organisms—as developed by Cairns-Smith (1985)—open wide the possibilties for replicating, information-transferring, and even (elementary) photosynthesizing kaolinite crystals (for example) to redefine entirely the extent of the HZ.

**The Astrobiology Versus Xenobiology Debate**

The most profound challenge to the HZ concept comes from the advocates of xenobiology, who assert that life could exist in extremely unEarth-like habitats; that it need not resemble life as we know it, and that it might have drastically different ecophysiological requirements. Many scientists may prefer not to engage with a concept which involves purely hypothetical life-forms and which has been best explored through science fiction, but a robust epistemology of planetary habitability should be able to accommodate discussion of radical hypothetical possibilities.

Xenobiology apostles Cohen & Stewart (2001; 2002) have contended that astrobiology (the name under which the study of life beyond the Earth is generally pursued) is a science of self-inflicted limitations, that it involves a lack of imagination and a failure of nerve.

A couple of quotes from their book "*Evolving the Alien*" summarise their position.

Cohen & Stewart (2002, p. 234): "'*Creatures' made from magnetic vortices in stellar plasmas could be at least as complex as any lifeform on Earth, with topological linkages of vortices taking the place of linear topology of DNA. And they would only survive inside a star: that's the best place for them to live. (Even more so, what price 'habitable zones'?)*"

Cohen & Stewart (2002, p. 338): "*Planet Earth is one tiny, unrepresentative lump of rock in an exceedingly ordinary part of a vast, incomprehensible universe.*"

It would not be fair to dismiss all astrobiology as Earth parochialism, however, since some papers appearing in peer-reviewed astrobiological journals have explored xenobiology. Bains (2004) discussed the possibility of biological solvents other than water, and Schulz-Makuch *et al.* (2004,) encouraged by the work of Sattler *et al.* (2001), who described not merely viable, but actively reproducing bacteria in water droplets in high altitude clouds, Schulz-Makuch *et al.* (2004, p. 11) contended that: "'*Having originated in a hot proto-ocean or been brought in by meteorites from Earth (or Mars), early life on Venus could have adapted to a dry, acidic atmospheric niche as the warming planet lost its oceans. The greatest obstacle for the survival of any organism in this niche may be high doses of ultraviolet (UV) radiation. . . . such an organism may utilize sulphur allotropes present in the venusian atmosphere, particularly $S_8$, as a UV sunscreen, as an energy-converting pigment, or as a means for converting UV light to lower frequencies that can be used for photosynthesis.*"

Bains (2004) envisaged a zone where liquid $H_2SO_4$ might be found, which might extend Sun-ward of 1.0 AU around the present day Sun, and liquid $CH_4$ as well as liquid $N_2$ zones, which might extend outwards. These zones somewhat overlapped, but their existence need not be problematic (and until it can be shown that liquids other than water can serve as biological solvents, they are purely hypothetical).

Lorenz *et al.* (1997, p. 2,905), discussing the future of Titan, envisaged: "*Similar environments, with approximately 200 K water-ammonia oceans warmed by methane greenhouses under red stars, are an alternative to the approximately 300 K water-$CO_2$ environments considered the classic 'habitable' planet.*"

Whatever one's opinion about the viability of xeno-life, the stumbling block posed by the xenobiology is epistemologically non-trivial. If the domain in environmental phase space occupied by life-as-we-know-it is immediately adjacent to or overlapping with other habitable regimes which might just as easily have appeared if the Earth had been hotter or colder, or had acquired a different complement of surface volatiles, then the notion of any kind of environmental window for life (of which the circumstellar HZ is an obvious example) is scientifically meaningless. If this is the case, then, our supposed environmental window for life would be nothing more than a list of conditions to which life on Earth has become adapted. We would have fallen into the trap of holding up a mirror to life on Earth and calling it an image of life elsewhere.

It is possible that life on Earth occupies part of a continuum of opportunities for life, in relation to which life may evolve with a continuum of possible biochemical and anatomical adaptations. If so, the apparent environmental window for Earth-type life may be overlapped by other windows, and they in turn overlapped by others, with habitable regimes (involving appropriate organisms and modes of ecosphere function with their own finite ranges in terms of environmental parameters) extending into the distant reaches of the phase space of the possible.

Again, life and ecospheres may inhabit a probabilistic landscape in which they generally occupy islands in phase space which are separated by regions in which the opportunities for life have negligible probabilities. Bio-engineered life and ecospheres might extend the boundaries of these islands, or even infill domains between islands.

On the other hand, we must recognise that, at the present day, there cannot be any guarantee that xenolife is viable. It may be that life on Earth occupies the only available island in environmental phase space. In favour of that conclusion is the apparent absence (to date) of macro-organisms on any other Solar System body than the Earth, and the fact that high-biomass forest ecosystems have not managed to occupy certain Earth environments—notably ice caps and deserts—or to develop floating masses on the oceans or in the atmosphere.

Here is a potentially fruitful field of research for biologists, mathematicians and others that is awaiting attention. It lies at the heart of the philosophical discussion about the HZ.

The problem which xenolife poses for present purposes is not that of adjudicating proposed life processes in terms of their viability, but of how to accommodate such biota within a consistent framework. It cannot, for obvious reasons, be accommodated within the classic HZ, and the term HZ could only be useful if it were accompanied in each instance by a reference to the precise circumstances of habitability under consideration. Those circumstances would include, inevitably, the structure and function of ecospheres, and so it is more convenient, we contend, to discuss possibilities from the outset in the explicit context of Ecodynamic Domains.

**Multiple Ecodynamic Regimes Within the Classic HZ**

The concept of Ecodynamic Domains comes into its own for investigations of the multiple regimes which may exist for geophysically similar Earth-like planets orbiting within the classic HZ, or even receiving the same amount of insolation.

Tajika (2008), who considered the Earth, by virtue of its significant water cover, to be an "*ocean planet*", observed that (p. L53): "*The region around a star which satisfies the condition for the presence of liquid water is termed the habitable zone (HZ) (Hart 1979; Kasting et al. 1993). An ocean planet is, however, known to have multiple climate modes, including an ice-free state, a partially ice-covered state, and a globally ice-covered state . . . This means that the ocean planet could be globally ice-covered, even if it is within the HZ.*"

There is widespread support amongst geologists (with some dissenters) for interpreting the Precambrian record in terms of episodes of glaciation. Young *et al.* (1998) discussed examples of dropstones from the 2.9 Gyr old Mozaan Group of the late Archaean (3.8 to 2.5 Gyr ago). More startling is evidence for *low latitude glaciations*, which advocates of the "*Snowball Earth*" hypothesis (for example, Hoffman *et al.*, 1998; Kirschvink *et al.*, 1992; 1998; 2000) interpret in terms of a model with the oceans frozen over to depths of maybe 2.0 km. An entirely ice covered Earth would experience very low weathering rates, so outgassed $CO_2$ could build up in the atmosphere until catastrophic de-glaciation occurred and $T_{surf}$ soared. A global palaeomagnetic compilation of

Earth's entire basin scale evaporite record by Evans (2006) indicated that for most of the past 2.0 Gyr, measured palaeo-magnetic inclinations were compatible with the Earth having a low obliquity as today (rather than a problematic high obliquity) and that there was also, as today, a geocentric-axial-dipole magnetic field. It was concluded (Evans, 2006, p. 51) that: "*the snowball Earth hypothesis accordingly remains the most viable model for low-latitude Proterozoic ice ages.*" Notwithstanding, climate simulations by Hyde *et al.* (2000) found difficulties in obtaining completely ice-covered oceans, except for $pCO_2$ around or lower than PAL. Peltier *et al.* (2007, p. 813) concluded that: "*drawdown of atmospheric oxygen into the ocean, as surface temperatures decline, operates so as to increase the rate of remineralization of a massive pool of dissolved organic carbon. This leads directly to an increase of atmospheric carbon dioxide, enhanced greenhouse warming of the surface of the Earth, and the prevention of a snowball state.*"

Heath (1990a, 1990b) emphasised that the geological record and models for future environmental trends defined between them a window in Earth history during which a metabiota could thrive. This approach exploited the fact that the carbonate-silicate cycle provided a potential relationship between several factors, including stellar evolution, photospheric effective temperature, insolation, planetary obliquity, geophysical-petrochemical evolution, $T_{surf}$, and the distribution of climatic zones, continental distribution, $pCO_2$ and photosynthetic productivity.

Credit for laying the foundations of this approach belongs to Walker *et al.* (1981), who postulated the carbonate-silicate cycle, and to Lovelock & Whitfield (1982), who investigated the fate of photosynthetic production in response to the long-term downward adjustment of $pCO_2$ as insolation rises during the Sun's main sequence evolution. The latter authors feared that photosynthetic production would cease ~ 0.1 Gyr from now when just 150 ppm $CO_2$ remained in the atmosphere. The latter figure corresponded to the $CO_2$ compensation point, at which photosynthetic carbon fixation would exactly balance losses through respiration. We stress that this must not be confused with the $CO_2$ value at which healthy metabolism ceases; with zero net $\overline{C}$ fixation available, a plant could not repair itself, produce seeds, or even grow.

Heath (1990a, p. 1, Meeting: *Life & Death of the Earth*, March 31, 1990, London) wrote: "*rising solar luminosity and secular cooling of the Earth's interior, accompanied by continental growth, drive environmental changes that define a unique window for the proliferation of the metabiota. This will be much shorter (< 1.0 Gyr) than the Sun's main sequence life-span (10 Gyr).*"

Continuing petrogenesis of continental rock, with associated increase in the area of emergent continent, would have promoted $CO_2$ drawdown, and provided geological environments for the long-term residence of both shelf carbonates, the sink for atmospheric $CO_2$, and for photosynthetically-produced organic C (thus driving up $pO_2$ towards levels that could support the respiratory demands of higher plants and animals).

Moreover (Worsely & Nance, 1989), as geothermal flux declined, the hydrothermal exchange temperature at hot water vents at the Mid Ocean Ridge system would also decline, which would have reduced the ability of Fe and Mn oxyhydroxides to scavenge the nutrient P from sea water. This could have boosted photosynthetic production, organic C burial and thus atmospheric and oceanic oxygenation.

Of interest are also the papers by Des Marais *et al.* (1992), who argued that rifting in continental crust was particularly important in providing basinal environments for the accumulation of sediments containing organic carbon, and Moores (1993), who argued from observation of preserved ophiolite sections that oceanic crust became distinctly thinner at ~ 1.0 Gyr ago as a result of declining geothermal flux and reduced volcanism; this must have caused sea levels to fall with a concomitant substantial increase in the area of emergent continental rock.

It was evident that much would depend upon details of planetary history, and that a multiplicity of ecodynamic regimes could exist for any level of insolation. Heath (1992, Greenwich College internal document) observed: "*the timing of the differentiation and growth of continental crust, total and emergent continental areas, and the rate of $CO_2$ outgassing should vary from one otherwise Earth-like planet*

*to another"*. This paper stated that the HZ *"cannot be defined with regard to a broad class of hypothetical Earth-like planets. Rather it is necessary to specify a precise planetary condition."*

Heath (1992) re-examined the width of the HZ for human habitation with reference to ecophysiological constraints for Earth-type plant life, as well. At that stage, following Dole (1964) this zone was designated the *"ecosphere."* The reader is reminded that we apply the term here in another sense, namely that sense preferred by ecologists—namely, a planetary-scale ecosystem. However, it is interesting to note that for Dole (1964), who focussed on the possibility of planets naturally available for human colonisation—and whose *"ecosphere"* was effectively a circumstellar human habitable zone—the *"ecosphere"* was to be defined as a zone where insolation levels were (p. 64) *"compatible with the origin, evolution to complex forms, and continuous existence of land life and surface conditions suitable for human beings, along with the ecological complex on which they depend"*. Terminological issues aside, Dole's recognition of an ecological complex pre-figured (and partly inspired) the approach which we were later to adopt.

Heath (1992) used models of $T_{surf}$ and $pCO_2$, based on the carbonate-silicate cycle which had been developed by Marshall *et al.* (1988) for polar, mid-latitude and equatorial continental configurations and lower past solar luminosity. The interesting implication for the ability of Earth-like planets to support complex life was that significant differences in of $T_{surf}$ and $pCO_2$ pertained for different continental configurations.

The results of Marshall *et al.* (1988) for lower insolation were re-applied by Heath (1992) for the appropriate range of distances from the present day Sun. The outer edge was deemed to be defined by the distance at which $pCO_2$ exceeded dangerous levels for long-term human respiration (5,000 ppm, according to Masterlerz, 1977; 9,000 ppm, according to Dole, 1964). Marshall *et al.* (1988) had not modelled higher insolation than that of the present day Earth, so the inner edge was taken as that which $pCO_2$ fell below 150 ppm, as in the model of Lovelock & Whitfield (1982). Subsequently, Worsley & Nance (1989) had pointed out that ≈ 150 ppm applied to plants with a type of metabolism known as $C_3$ (the first stable product of photosynthesis is a three carbon molecule, whilst for plants with $C_4$ metabolism the first stable product of photosynthesis is a four carbon molecule) can have very much lower $CO_2$ compensation points. (Note that most forest trees use $C_3$ metabolism).

This human-habitable zone ran from perhaps 0.976 – 0.995 AU out beyond 1.06 AU ($pCO_2$ rose very rapidly for all continental configurations beyond 1.12 AU. This principle was then applied to stars of solar composition and masses 0.88 $M_s$, 1.00 $M_s$ and 1.12 $M_s$, from the evolutionary models of Demarque *et al.* (1986), and from the same authors for the main-sequence G2 dwarf alpha Centauri A (1.085 $M_s$) for metallicity of 0.02 (approximating the solar value) and 0.04. ("Metallicity" is astronomical parlance for the proportion of elements heavier than hydrogen and helium.) The higher the metallicity for a star of given mass and proportion of hydrogen and helium, the lower will be its luminosity. It was noted that climatic models by Gérard *et al.* (1992), the Earth may have been prone to global freezing events until insolation reached 0.908 - 0.969 of the current value. The higher value would have pertained in the early Palaeozoic, actually after the first land plants had appeared, illustrating how the width of the forest-habitable and human habitable zone are subject to model-dependent arguments - or to happenstances of planetary history.

It was concluded (p. 3) that *"Recognizing that the timing of the differentiation and growth of continental crust, total and emergent continental areas* [controlling the amount of crust available to promote $CO_2$ drawdown through weathering], *and the rate of $CO_2$ outgassing should vary from one otherwise Earth-like planet to another, it must be concluded that the ecosphere* [here used as human habitable zone] *cannot be defined with regard to a broad class of hypothetical Earth-like planets. Rather it is necessary to define a precise planetary condition."* Moreover, whilst work such as Wetherill (1991) had indicated that formation of planets of about an Earth-mass would be very common at distances of 0.8 to 1.3 AU from more or less Sun-like stars (this was before the discovery of giant planets close to their stars introduced complications into the picture), *"it cannot be assumed that a large proportion of these bodies would be suitable for human life and Earth's ecosystems."*

Subsequent work by Caldeira & Kasting (1992a) indicated that Marshall *et al.* (1988) had been too optimistic about how close ice sheets could advance towards the equator; according to Caldeira & Kasting (1992a) runaway glaciation would occur if ice advanced within 30º of the equator. It was concluded (Heath, 1992, p. 3) that "*Recognizing that the timing of the differentiation and growth of continental crust, total and emergent continental areas* [controlling the amount of crust available to promote $CO_2$ drawdown through weathering], *and the rate of $CO_2$ outgassing should vary from one otherwise Earth-like planet to another, it must be concluded that the ecosphere* [here used as the human habitable zone] *cannot be defined with regard to a broad class of hypothetical Earth-like planets. Rather it is necessary to define a precise planetary condition.*" Moreover, while work such as Wetherill (1991) had indicated that formation of planets of around an Earth-mass would be very common at distances of 0.8 to 1.3 AU from more or less Sun-like stars (this was before the discovery of giant planets close to their stars confused the picture), "*it cannot be assumed that a large proportion of these bodies would be suitable for human life and Earth's ecosystems.*"

Caldeira & Kasting (1992b) explored photosynthetic production in response to rising insolation. They noted that carbon dioxide was just a few ppm for $C_4$ plants, hence it might be rising temperatures and the progressive loss of Earth's oceans which terminated life on Earth, not $CO_2$ starvation. Caldeira & Kasting (1992b) predicted that as $T_{surf}$ warms above 60 to 70ºC, the $H_2O_{vap}$ mixing ratio in the stratosphere will increase markedly, and around 80ºC, it will become ~ 2.5%. Above this level, the loss of hydrogen atoms to space will be limited by the heating rate from Extreme Ultra-Violet (EUV) in insolation to ~ $6 \times 10^{11}$ atoms $cm^{-2} s^{-1}$, which will give a time-scale for ocean loss of ~ 1.0 Gyr. The oceans should have been lost by ~ 2.5 Gyr from now.

The predictions of this model for $T_{surf}$, $pCO_2$ and photosynthetic productivity were incorporated into the presentation from Heath (1994; 1996) to the *First International Conference on Circumstellar Habitable Zones*. A similar approach to habitability, also using the model by Caldeira & Kasting (1992b) as a point of departure, was evolved later and separately by another group.

Following this line of thought, von Bloh *et al.* (2009, p. 593) wrote: "*Another definition of habitability first introduced by Franck et al. (2000a, 2000b) is associated with the photosynthetic activity of the planet, which critically depends on the planetary atmospheric $CO_2$ concentration. This type of habitability is, thus, strongly influenced by the planetary geodynamics and encompassing climatological, biogeochemical, and geodynamical processes (''Integrated System Approach'').*" This approach likewise recognised that continuing formation of continental rocks would increase the land area and so adjust $pCO_2$ downwards by exposing a larger area for weathering.

The continued petrogenesis of continental crust as a consequence of continuing plate tectonics would tend to increase the area of continent available for weathering and hence accelerate $CO_2$ drawdown (Franck *et al.*, 2000b). It has been recognised for some time that biotic effects might prolong this process.

Moreover, the biota will not be passive in the process of $CO_2$ drawdown through weathering (Schwartzman & Volk, 1989). Schwartzman (1999) argues—contary to the concept of "*Snowball Earth*"—that the Precambrian Earth was too hot for advanced life, but that $T_{surf}$ was reduced by the biota itself.

The model of Lenton & von Bloh (2001) predicted a slower rise in temperature and reduction in $pCO_2$ than might otherwise be expected since (p. 1,715): "*life cools the Earth by amplifying the rate of silicate rock weathering and maintaining a low level of atmospheric $CO_2$. Recent studies indicate a much stronger biotic weathering effect than in models used to estimate the life span of the biosphere. . . . the resulting feedback lengthens the survival of complex life by delaying the loss of $CO_2$ from the atmosphere. The weathering biota can potentially maintain the Earth in a habitable state when otherwise it would be too hot for them. If so, catastrophic warming rather than gradual $CO_2$ starvation will terminate complex life.*"

In addition, von Bloh *et al.* (2007, pp. 1-2) specified an HZ for photosynthetic organisms: "*we adopt a definition of the HZ previously used by Franck et al. (2000a,b). Here habitability at all times does not just*

*depend on the parameters of the central star, but also on the properties of the planet. In particular, habitability is linked to the photosynthetic activity of the planet, which in turn depends on the planetary atmospheric $CO_2$ concentration together with the presence of liquid water, and is thus strongly influenced by the planetary dynamics. We call this definition the photosynthesis-sustaining habitable zone, pHZ. In principle, this leads to additional spatial and temporal limitations of habitability, as the pHZ (defined for a specific type of planet) becomes narrower with time due to the persistent decrease of the planetary atmospheric $CO_2$ concentration."*

This approach, developed through a succession of interesting and useful papers from workers at the Potsdam Institute for Climate Impact Research, was similar to ours in many ways, but not identical.

The major difference was that we were investigating complex ecosystems and forest biomass. Several other workers in this field have expressed more interest in photosynthetic production in general. As Chyba (1997, p. 201) pointed out: "*What it takes for a world to be habitable depends on who will be doing the inhabiting. Clearly, the requirements of humans differ from those of forests* [Heath, 1996], *which in turn are more stringent than those of say, green slime. Habitability for putative extraterrestrial technical civilizations may be the most interesting to consider. Yet since the history of surface life on Earth is largely the history of green slime, its needs must remain of special interest to exobiologists."*

Kasting *et al.* (1993) had defined the HZ in the following terms (p. 109): "*We ourselves are more interested in determining if life can evolve on other planets than we in colonizing them, so we will use the prevalence of liquid water as our habitability criterion. For ease of presentation, we will refer to the liquid-water region as the habitable zone, or "HZ," recognizing that not all planets in this region would make suitable homes for humans."*

Caldeira & Kasting (1992b) were probably correct in arguing that since some organisms, (for example, photosynthetic microbes) could function with just a few parts per million atmospheric $CO_2$, and since it might be difficult to eliminate such trace quantites from the atmosphere, it would be rising temperatures that would actually terminate photosynthesis in the Earth's far future, and water loss that eventually rendered any form of biosphere impossible as the solar output increases.

This, however, does not necessarily mean that *forests*, which contain most of the Earth's surface biomass, could function at low $pCO_2$. The reason is that forests require a substantial annual turnover in carbon. We may draw an analogy from the business world. Some sort of minimalist business might be run by a person in a third world country, living in a shack without services, who needs a profit of, say, just $ 10 a day to keep everything ticking over. This is the equivalent of a microscopic plant requiring a vanishingly small amount of $CO_2$. A tree is the equivalent of a skyscraper. A skyscraper, with its substantial demands in terms of maintenance of its water, electricity, telecommunications and waste products, needs a major financial investment simply to operate from day to day. The $CO_2$ compensation point, it should be remembered, is the point of zero growth and repair. A plant may survive such conditions in the laboratory, where its other needs have been supplied for the limited duration of an experiment. In the real world, however, a plant will need to grow and repair itself, so it must operate within a healthy margin between ambient $pCO_2$ and the compensation point. Forests involve great numbers of trees. Trees lose material through leaf loss (seasonal or ongoing), mechanical damage (as in gales breaking boughs) and other attrition, denoted by the term "*litterfall,*" and through herbivory, as well. Forests cannot function without substantial carbon throughput (Heath, 2002).

Assuming, somewhat arbitrarily, that biological productivity must be within an order of magnitude of that which supports forests today, the HZ for forests might stretch from about 0.97 AU to just 1.1 AU, if the most pessimistic interpretations about global freeze over are drawn from Caldeira & Kasting (1992a). However, since it had been long known that (leaving aside the possibility of dramatic obliquity changes) there was geological evidence for low level glaciations occurring on the Earth just ~ 0.6 Gyr ago (with ~ 0.95 the current solar insolation), and since—even though the Earth emerged into a warmer regime—the smothering of the continents with ice would have been inimical to forests, the outer margin of the forest-habitable zone might be located as close to home as 1.026 AU (an illustrative figure; we are not claiming that climatic modelling or modelling of biological responses is actually this accurate). The implication, even so, was that it was not necessarily possible to define a CHZ for forests.

Lenton & von Bloh (2001) provided cause for optimism about the future of photosynthetic production, but the prospects for forests were not so good. Data presented by Salisbury & Ross (1978) indicated that the productivity of *Acer* (maple) declines in an almost linear fashion with $p\text{CO}_2$, collapsing at 50 ppm. A plant with such a partial pressure requirement would become extinct through $CO_2$ starvation before the era of sharp temperature rise predicted by Lenton & von Bloh (2001).

We stress here two considerations which have not figured previously in these discussions. Firstly, the distinctions between $C_3$ and $C_4$ plants as regards partial pressure of carbon dioxide is a generalisation. Secondly, as has been demonstrated experimentally, responses to such factors as $p\text{CO}_2$, temperature and light are interdependent. Heath *et al.* (1967) showed, from work on a lettuce cultivar, that there is a fall in $p\text{CO}_2$ with increasing light intensity and also a fall in the light compensation point with increasing $p\text{CO}_2$. Green & Lange (1995) demonstrated how, in a C3 plant (the liverwort *Marchantia*), $p\text{CO}_2$ rose sharply from ~ 40 ppm at 5°C to ~ 170 ppm at 35°C ($pO_2$ = present value = 21 %). With the rising temperatures and falling $p\text{CO}_2$ predicted in the geostatic option of Lenton & von Bloh (2001), *Marchantia* would be unable to grow at all 0.6 Gyr from now.

Plant physiologists have quoted low values for $p\text{CO}_2$ for some $C_3$ plants. Bjorkman & Berry (1973) had cited 50 ppm. It should be noted that much lower $p\text{CO}_2$ values have been recorded experimentally for $C_3$ plants. Fox *et al.* (1986) demonstrated surprising values of $p\text{CO}_2 = \geq 7.7$ ppm for leaves of the $C_3$ plant celery (*Apium graveolans* L. "Giant Pascal"), which possesses nothing like Kranz anatomy, nor $C_4$ metabolism. Of course. despite their helpful metabolic and anatomical adaptations, $C_4$ plants also decline in productivity as $p\text{CO}_2$ falls.

Would ecosystems adjust to support smaller tree forms, or might they support a smaller number of more widely spaced large forms? At what point would trees be replaced by non-arboreal plants and would there be a catastrophic shift from forest to, say, grasslands at a critical $p\text{CO}_2$ and $T_{surf}$? Another possibility is the evolution, over the coming hundreds of millions of years, of rather different biology; we might envisage "*bubble forests*," which might contain and control $H_2O$, $CO_2$ and $O_2$, beneath transparent canopies through which they conducted photosynthesis. This possibility received attention in discussions after the 1994 *First International Conference on Circumstellar Habitable Zones*, although time prevented a formal presentation. Heath (1994) noted in his conference abstract (in Doyle, 1996, p. 493): "*In principle . . . biology could drive the equivalent of Earth's biogeochemical cycles on geologically inert worlds. Biological membranes, secretions and constructs interposed at the atmosphere-ocean, crust-ocean, and crust-atmosphere interfaces could control all aspects of gas exchange and weathering, and even modify planetary albedo. However, the more extreme the global environment to be overcome, the less likely it is that suitable biological systems could ever evolve.*"

Kasting *et al.* (1993) set the agenda for much of the research that followed. Any suggestion that there has been a hiatus since that publication would underestimate the impetus which this paper imparted to other workers.

From our perspective, one of the most important of their conclusions was that it reinforced our understanding that a planet at a given distance from its star might—depending upon the spectral quality of insolation and planetary characteristics—occur in a multiplicity of states, and furthermore, the margins of both the classic HZ and the most favourable zone for forests would vary accordingly.

Kasting *et al.* (1993) considered how the margins of the HZ might vary for planets of different mass: (p. 117): "*Near the inner edge of the HZ, and at low pCO₂ values near the outer edge, the increased greenhouse effect dominates; at high pCO₂ levels near the outer edge, the increased albedo dominates. The maximum greenhouse limit changes very little because it represents a trade-off between the two competing*

*effects. Since the inner edge moves inward while at least one estimate of the outer edge remains fixed we conclude that, for a given surface pressure, large planets may have somewhat wider HZs than do small ones."* They considered also how the margins of the HZ would vary with the effective photospheric temperature ($T_{eff}$) of parent stars. They adopted $T_{eff}$ = 3,700 K for an M0 star (0.5 $M_s$, 0.06 $L_s$); $T_{eff}$ = 5,700 K for a G2 star (1.0 $M_s$ 1.0 $L_s$); $T_{eff}$ = 7,200 K for an F0 star (1.3 $M_s$ 4.3 $L_s$). These $T_{eff}$s and luminosities were at selected times in their main sequence histories. Because hotter stars produce light biased towards shorter wavelengths—and because Rayleigh scattering of light varies according to the inverse fourth power of the albedo—for an Earth-like planet, all else being equal, the atmospheric temperature would be higher the hotter the star.

Responding to the models for the HZ proposed by Kasting *et al.* (1993), Heath (1993, pp. 1-2) observed that *"The Archaean Earth, with less insolation than today, may have enjoyed high-$CO_2$ greenhouse conditions, thanks to greater outgassing and less continental area available for weathering, but the latter would not favour a large land biomass. A planet with more land than Earth, but lower outgassing, might suffer major $CO_2$ drawdown near 1.0 A.U.. The range of geophysics, geochemistry, and geological activity on basically Earth-like planets remains unknown.*

*Moreover, the geological record attests to significant fluctuations in climate and atmospheric composition, not smooth evolution as solar output has increased, so that a planet might suffer loss of biomass within an apparently favourable insolation range. The distribution of seas, land, and mountain ranges, lakes and long term changes in outgassing rate might become of key importance where other factors are critical (I thank A. F. Machiraju for drawing to my attention Yemane, 1993, which argues that extensive lakes prevented the harsh climate otherwise modelled for Late Permian Gondwanaland)."*

A further complication was introduced by Laskar & Robutel (1993) and Laksar *et al.* (1993), who demonstrated that the Earth, if rotating at its present rate, and without a large Moon at the right distance, would be subject to chaotic changes in obliquity. Over a period of around 20 million years, its tilt could vary unpredictably from a seasonless zero degrees to around ninety degrees, when the Sun would pass overhead at the poles once each year, and seasons would be extreme.

The implications of obliquity and continental distribution were investigated further by Williams & Kasting (1997), which investigated the effects of changing obliquity. Darren Williams kindly carried out a few more simulations for us which enabled us to investigate the heliocentric distances at which $CO_2$ drawdown would undermine the maintenance of forest biomass for different land configurations and axial tilts (Heath & Doyle, 2000; Heath *et al.*, 2000).

On a planet receiving < 1.0 current solar insolation, predicted equilibria for $pCO_2$ are high. Williams & Kasting (1997) calculated that for an Earth-like planet at 1.4 AU, receiving 0.51 solar insolation that $pCO_2$ would be ~ 2.0 bars. It must be asked whether higher plants would fail to evolve, or whether other environmental factors, such as competition between species to exploit reduced light levels would drive evolution towards complex photosynthetic life forms. Robinson (1991) argued that evolutionary changes in land plants may have been prompted by a low term trend in Earth history for falling $pCO_2$ (according to Berner, 1992, from ~ 15 PAL in the Middle Ordovician to ~ 8 PAL in the Early Devonian), as plants had to become more efficient at gaseous exchange whilst minimising simultaneous water loss. In the standard model of Kasting and co-workers, 5,000 ppm $CO_2$ (the safe upper limit for humans and ~ 18 PAL) would occur around 1.05 AU. In our minds, the discussion in Robinson (1991) raises the question of whether plants on Earth-like planets with < 0.90 current insolation—where high $pCO_2$, might mean that there was significantly less evolutionary pressure for development of higher plant forms—would tend to remain in more rudimentary forms.

Further complications, with the potential to constrain the boundaries of ecodynamic domains, would be introduced in a detailed discussion of alternative structures which might exist in other planetary systems. We defer that task to a future paper.

Values of insolation and geothermal flux for critical transitions between ecosphere boundaries will continue to be revised as models improve. Such work will provide the context within which ecodynamic perspectives can develop and evolve.